\documentclass[aps,prb,reprint,superscriptaddress,twocolumn,showpacs,floatfix]{revtex4-1}
\setcitestyle{square,numbers}
\usepackage{graphicx}  
\usepackage{bm}        
\usepackage{amssymb}   
\usepackage{amsmath}
\usepackage{color}
\usepackage{hyperref}

\begin{document}

\title{Grand canonical Peierls transition in In/Si(111)}
\author{Eric Jeckelmann}
\email[E-mail: ]{eric.jeckelmann@itp.uni-hannover.de}
\affiliation{Institut f\"{u}r Theoretische Physik, Leibniz Universit\"{a}t Hannover, Appelstr.~2, 30167 Hannover, Germany}
\author{Simone Sanna}
\author{Wolf Gero Schmidt}
\affiliation{Lehrstuhl f\"ur Theoretische Materialphysik, Universit\"at Paderborn, D-33095 Paderborn, Germany}
\author{Eugen Speiser}
\author{Norbert Esser}
\affiliation{Leibniz-Institut f\"ur Analytische Wissenschaften, ISAS e.V., Schwarzschildstrasse~8, D-12489 Berlin, Germany}

\date{\today}

\begin{abstract}
Starting from a Su-Schrieffer-Heeger-like model inferred from
first-principles simulations, we show that the metal-insulator transition
in In/Si(111) is a first-order grand canonical Peierls transition in which
the substrate acts as an electron reservoir for the wires. This model explains
naturally the existence of a metastable metallic phase over a wide temperature
range below the critical temperature and the sensitivity of the transition
to doping. Raman scattering experiments corroborate the softening
of the two Peierls deformation modes close to the transition.
\end{abstract}


\maketitle

A Peierls-like transition in indium wires on the Si(111) surface was 
first reported 16 years ago~\cite{yeom99}.
Since then this transition has been studied 
extensively~\cite{sni10,cho01,flei03,tani04,ahn04,tsay05,gonz05,gonz06,lope06,riik06,flei07,gonz09,wipp10,spei10,hatt11,schm12,wall12,kim13,klas14,zhan14} 
both experimentally and theoretically.
The occurrence of both a metal-insulator transition around $T_c=130 K$ and a structural transition
of the In wires from a $4 \times 1$ structure at room temperature to a $8 \times 2$ 
structure at low temperature are well established.
Yet, the nature of the transition is still poorly understood and the relevance of the 
Peierls theory remains controversial~\cite{gonz05,gonz06,riik06,flei07,wipp10,kim13,kim16}.

The generic theory of Peierls systems is essentially based on effective models for the low-energy degrees
of freedom in purely one-dimensional (1D) or strongly anisotropic three-dimensional (3D) crystals,
such as the Ginzburg-Landau theory of 1D charge-density waves (CDW)~\cite{gruener00} or the Su-Schrieffer-Heeger (SSH)
model for conjugated polymers~\cite{su79,su80,heeg88,baeriswyl92,barford}.
Hitherto it has been used without adaptation to discuss the relevance of the Peierls physics for experiments and first-principles 
simulations in In/Si(111). 
Thus a fundamental issue with previous 
interpretations based on these generic theories
is that they do not consider
how the 3D substrate affects the Peierls physics in a 1D atomic wire.

In this Rapid Communication, we investigate the phase transition in In/Si(111) theoretically using first-principles simulations and
1D model calculations, and experimentally with Raman spectroscopy.
We show that it can be interpreted as a grand canonical Peierls transition,
in which the substrate acts as a charge reservoir for the wire subsystem.
The two Peierls distortion modes are essentially made of shear and rotary modes.
The main difference with the usual (i.e., canonical) Peierls theory is that in the grand canonical theory the high-temperature phase can remain 
thermodynamically metastable below the critical temperature $T_c$ and that the phase transition can
become first order. This agrees with the interpretation of recent experiments and first-principles simulations
in In/Si(111)~\cite{hatt11,wall12,schm12,klas14,zhan14}.

\begin{figure}[tb]
\includegraphics[width=0.44\textwidth]{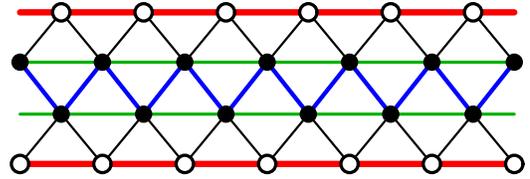}
\caption{\label{fig:lattice} (Color online) 1D lattice model for an indium wire
in the uniform configuration.
Open and full circles represent outer and inner In atoms, respectively.
The line widths are proportional to the hopping terms $t_{ij}$.
The blue and red bonds define the central zigzag chain and the two outer linear chains, respectively.
}
\end{figure}

First, we construct an effective 1D model for In/Si(111) in the spirit of the SSH model~\cite{su79,su80,heeg88,baeriswyl92}.
Our goal is a qualitative description of the phenomena with reasonable order of magnitudes for physical
quantities because we think that a quantitative description of this complex material can only be achieved with
first-principles simulations~\cite{suppmat}.
For the same reason, we neglect correlation effects~\cite{baeriswyl92,barford,jeck94,jeck98b}.
The accepted structural model for the uniform phase (i.e., the $4 \times 1$ phase) consists in parallel pairs of zigzag indium chains~\cite{bunk99,naka01}.
We consider a single wire made of four parallel chains of indium atoms arranged on a triangular lattice as shown in Fig.~\ref{fig:lattice}.
One (Wannier) orbital per indium atom is taken into account yielding four bands in the uniform phase.
Density-functional theory (DFT) calculations actually show four bands corresponding to indium-related surface states~\cite{gers14}.
Other electronic degrees of freedom, e.g. in the substrate, are not considered explicitly.

\begin{figure}[tb]
\includegraphics[width=0.49\textwidth]{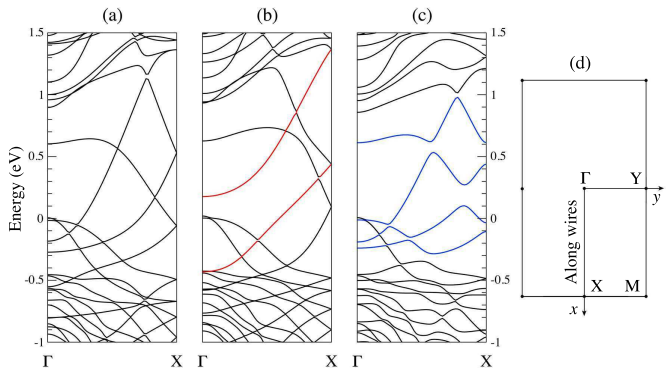}
\caption{\label{fig:bands} (Color online)
DFT-LDA electronic band structure of In/Si(111) (a) in the $4 \times 1$ phase,
(b) after a shear distortion, and (c) after a rotary distortion in
the surface 
Brillouin zone of the $4 \times 2$ configuration 
shown in (d).
Gaps open (b) at $\Gamma$ between
two red bands and (c) close to $X$ between four blue bands.
}
\end{figure}

We use a tight-binding Hamiltonian model for the electronic degrees of freedom and assume
that the only relevant hopping terms are between nearest-neighbor sites, i.e., 
\begin{equation}
\label{eq:hamiltonian}
H = \sum_{i,\sigma} \epsilon_{i} c^{\dag}_{i\sigma} c^{\phantom{\dag}}_{i\sigma} 
- \sum_{\langle i,j\rangle, \sigma} t_{ij} \left ( c^{\dag}_{i\sigma}
c^{\phantom{\dag}}_{j\sigma} + c^{\dag}_{j\sigma} c^{\phantom{\dag}}_{i\sigma} \right )
\end{equation}
where the indices $i,j$ number the indium atoms, $\sigma=\uparrow,\downarrow$
designs the electron spin, the second sum runs over every pair
$\langle i,j\rangle$ of nearest-neighbor sites, and
the operator $c^{\dag}_{i\sigma}$ ($c^{\phantom{\dag}}_{i\sigma}$) creates
(annihilates) an electron with spin $\sigma$ on site $j$.  
In the uniform phase the Hamiltonian is translationally invariant 
and the single-electron dispersions can be calculated analytically~\cite{suppmat}.
Thus we can determine parameters $\epsilon_{i}$ and $t_{ij}$ to mimic the 
DFT band structure~\cite{schm12,gers14} shown in Fig.~\ref{fig:bands}(a). We obtain three metallic bands
and one full band if we assume that the 1D system is close to half filling (i.e., one electron per orbital on average).

The strength of the hopping terms $t_{ij}$ is shown in Fig.~\ref{fig:lattice}.
Clearly, the apparent structures are a central zigzag chain and two outer linear chains.
The bond order (electronic density in the bonds between atoms) exhibits a similar structure~\cite{suppmat}.
This is quite different from the usual representation of the $4 \times 1$ configuration by two zigzag chains.
Our effective 1D model focuses on the metallic bands and thus reveals 
the bonds responsible for the Peierls instability.

In the hexamer structural model for the low-temperature phase,
the deformation from the uniform to the dimerized (i.e., $8\times 2$ or $4 \times 2$) phase
corresponds essentially to the superposition of two rotary and one shear modes~\cite{gonz05,gonz06,riik06,spei10,schm12}.
Therefore, we investigate the changes in the lattice structure, electronic band structure, and electronic density
caused by each mode separately using
first-principles frozen-phonon and deformation-potential calculations based on
DFT within the local density approximation (LDA). 
The technical details correspond to earlier calculations by some of the present authors~\cite{wipp10,schm12}.
A very recent hybrid DFT calculation~\cite{kim16} largely agrees with the DFT-LDA results presented here.
We use 
distortion amplitudes close to the ones necessary to transform the zigzag structure into the hexagon structure.
The predicted vibration modes agree well with Raman spectroscopy measurements presented here and in
previous works~\cite{flei03,flei07,spei10}.

\begin{figure}[tb]
\includegraphics[width=0.44\textwidth]{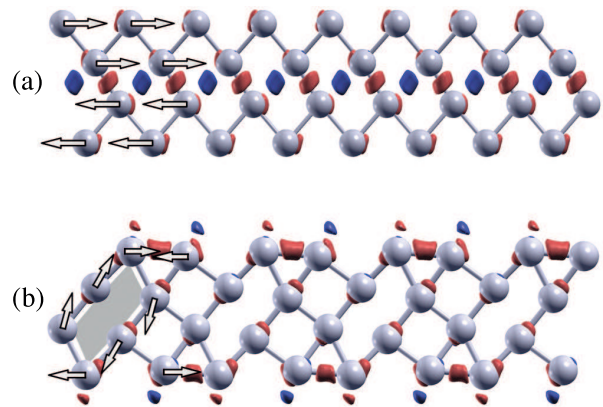}
\caption{\label{fig:density} (Color online)
Changes in the DFT-LDA electronic densities (red for an increase, blue for a reduction) 
with respect to the $4 \times 1$ phase caused by (a) a shear distortion and (b) a rotary distortion.
The isosurfaces for density changes $\pm 0.02$ e\AA$^{-3}$ are shown.
Arrows show the atom displacements for both distortion modes
}
\end{figure}

This study reveals on the one hand that the main effects of the shear distortion are to dimerize the central zigzag chain,
as shown by the alternating density  and bond lengths between inner In atoms in Fig.~\ref{fig:density}(a),
and to open or enlarge a gap between two metallic bands close to the $\Gamma$ point as seen in Fig.~\ref{fig:bands}(b).
On the other hand, the main effects of the rotary modes are to dimerize the outer chains, as shown by the alternating density and bond lengths between 
outer atoms in Fig.~\ref{fig:density}(b), 
and to open a gap between two metallic bands close to the $X$ point, as seen in Fig.~\ref{fig:bands}(c).
These results confirm the central role of the structures seen in Fig.~\ref{fig:lattice} (i.e., one inner zigzag chain and two outer
linear chains) in the transition of In/Si(111).
Moreover, the negligible length and density variation for the bonds between inner and outer indium atoms in first-principles calculations, 
both for shear and rotation distortions, confirm that they are very strong covalent bonds and do not play any direct 
role in the transition.

The SSH model~\cite{su79,su80,heeg88,baeriswyl92} 
is the standard model for the CDW on bonds caused by a Peierls distortion
seen in Fig.~\ref{fig:density}.
The bond length changes determined with first-principles methods can also be used to determine the hopping terms of the 1D  model~(\ref{eq:hamiltonian})
for distorted lattice configurations.
For this purpose, we assume that the hopping term between two orbitals $i$ and $j$ depends only on the distance
$d_{ij}$ between both atoms and choose the exponential form~\cite{long59,sale60}
\begin{equation}
\label{eq:coupling}
t_{ij}\left ( d_{ij} \right ) = t_{ij} \exp \left [ -\alpha_{ij} \left ( d_{ij} - d^0_{ij} \right )  \right ]
\end{equation}
where $t_{ij}$ and $d^0_{ij}$ are the hopping terms and bond lengths in the uniform configuration.
Using reasonable values for the electron-lattice couplings $t_{ij}\alpha_{ij}$ (i.e., $\alpha^{-1}_{ij}$ is of the order
of the covalent radius of an In atom),
we find a qualitative agreement between first-principles and 1D model predictions
for the changes in the band structure and density caused by shear and rotary modes~\cite{suppmat}.

The mechanism of the Peierls transition can be understood even better by focusing on the main features of the 1D model.
Keeping only the most important hopping terms (thick lines in Fig.~\ref{fig:lattice}) and couplings to lattice distortions,
the 1D model decouples into three independent chains with SSH-type Hamiltonians~\cite{su79,su80,heeg88,baeriswyl92} and
electron-lattice couplings~(\ref{eq:coupling}):
the inner zigzag chain which couples only to the shear mode and two identical outer linear chains which couple only to
one of the two rotary modes each.
To complete the SSH-type Hamiltonians we add an elastic potential energy for the lattice deformation.
The free energy of each chain ($l=1,2,3$) is then given by
\begin{equation}
\label{eq:potential}
F_{l}(x_{l})= F^{e}_{l}(x_{l}) + \frac{K_{l}}{2}  x^2_{l}
\end{equation}
where $ F^{e}_{l}$ is the electronic free energy~\cite{gruener00}.
Within this mean-field and semi-classical approach, the stable configurations are given by the minima of the total free energy 
$F=\sum_{l} F_{l}$ of the 1D model with respect to the amplitudes $x_{l}$ of the three independent distortion modes.
The bare elastic constants $K_{l}$ can be estimated using the distortion amplitudes $x_l$
necessary to form the hexamer structure in first-principles calculations~\cite{suppmat}.

This generalization of the SSH  model includes more degrees of freedom than the generalized SSH model used very recently to investigate chiral solitons
in indium wires~\cite{cheo15}.
Yet the model of Ref.~\cite{cheo15} corresponds essentially to the restriction of our model to outer chains and rotary distortions.
Furthermore, the model parameters found in Ref.~\cite{cheo15} also agree quantitatively with our parameters for outer chains and rotary distortions~\cite{suppmat}. 
In Ref.~\cite{kim16} Kim and Cho compare their DFT results to the two-chain SSH model of Ref.~\cite{cheo15} and
conclude that the transition in In/Si(111) is not a Peierls transition. However, their DFT results
seem to agree largely with our three-chain SSH model and thus support the Peierls transition scenario 
presented here. 

We can now analyze the 1D model in the mean-field approximation using known 
results for one-band/one-mode SSH-type models~\cite{su79,su80,heeg88,baeriswyl92}.
At half filling the outer chains have Fermi wave number $k_F=\pi/2$ and thus are unstable
with respect to rotary distortions with the nesting wave number $Q = 2 k_F  = \pi$
[corresponding to the $X$ point of the Brillouin zone in the $4\times 1$ configuration of In/Si(111)].
As the zigzag chain has two orbitals per unit cell, its Fermi wave number is $k_F=\pi$
and thus it is unstable against a shear distortion with the nesting wave number $Q = 2k_F = 2\pi$
(corresponding to the $\Gamma$ point).
Therefore, if the system is exactly half filled, the twofold degenerate ground state of each chain is a band insulator
with a dimerized lattice structure. 
The corresponding theoretical collective vibrational modes agree with the Raman spectroscopy results presented below.

This corresponds to an eightfold degenerate and insulating phase in the full 1D model.
The neglected couplings between the three chains reduce
the Peierls deformation modes to two linear combinations of the shear and rotary modes and
the degeneracy to four states corresponding to the four hexamer structures of the $4 \times 2$ phase.
The Peierls gap in the electronic band of the inner chain is at $k=0$
while Peierls gaps for the outer chains are at $k=\pi/2$ (i.e., the $X$-point
of the $4 \times 2$ configuration).
Typically, the electronic gap of the full 1D model is indirect and smaller than the Peierls gaps.
Thus there is no obvious relation between critical temperature and electronic gap
in this many-band Peierls system.
The structural transition to the high-temperature uniform phase is continuous but may 
exhibit distinct critical temperatures for shear and rotary modes. The metal-insulator transition occurs
at the lowest one.

This conventional Peierls scenario assumes a fixed band filling. 
The low-temperature insulating electronic structures found in DFT computations~\cite{gonz05,schm12,gers14}
correspond to half filling in the 1D model~(\ref{eq:hamiltonian}).  
However, for substrate-stabilized atomic chains, the electron chemical potential $\mu$ is determined by the substrate
and may be modified by temperature and adatoms~\cite{shim09,mori10,schm11}.
Therefore, we must investigate the 1D model in the grand canonical ensemble
with $\mu$ set by an external electron reservoir, i.e., the rest of the In/Si(111) system.
Focusing again on the decoupled 1D model, the free energies~(\ref{eq:potential}) are replaced by corresponding grand canonical potentials
$\phi_{l}$ and $\phi$.

\begin{figure}[tb]
\includegraphics[width=0.46\textwidth]{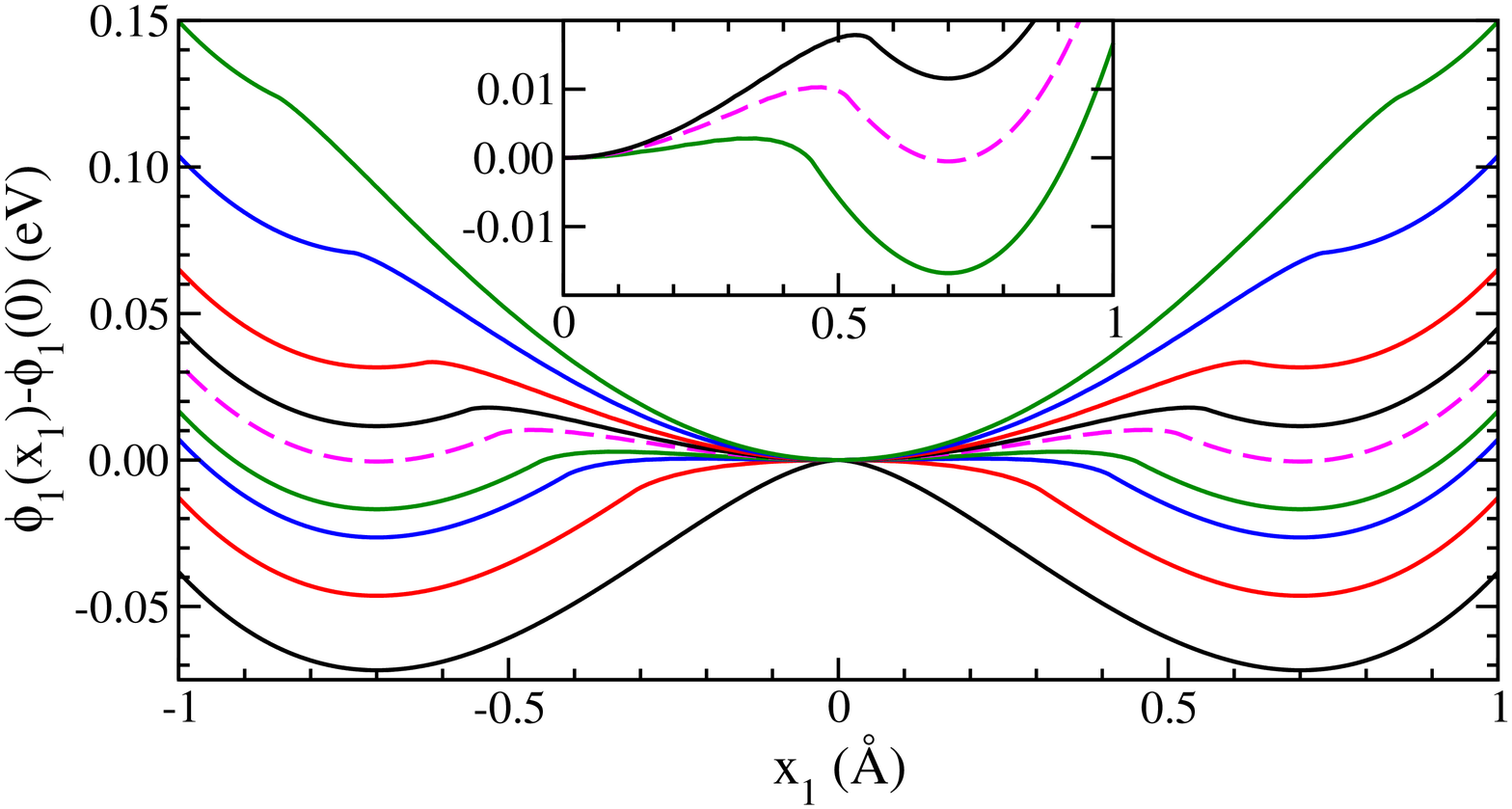}
\caption{\label{fig:potential} (Color online)
Grand canonical potential $\phi_1$ of the inner zigzag chain as a function of the amplitude of the shear 
distortion $x_1$ at low temperature 
for several values of the chemical potential $\mu$ from the middle of the gap (bottom) to the band edge (top).
The dashed line corresponds to the critical $\mu_c$ between dimerized and uniform phases.
Inset: Enlarged view close to $\mu_c$. 
The shift of $\mu$ between upper and lower bands corresponds to 7\% of the Peierls gap.
}
\end{figure}

We find that the grand canonical Peierls physics is much richer than the canonical one. 
Figure~\ref{fig:potential} shows the grand canonical potential of the inner chain at very low temperature as a function of the distortion amplitude
for several values of $\mu$. If $\mu$ lies at or close to the middle
of the Peierls gap, we see the usual double well indicating a stable and doubly-degenerate dimerized state.
When $\mu$ deviates slightly from the middle of the gap, a local minimum appears at $x=0$ indicating that the uniform state is metastable.
This case agrees qualitatively with the energetics of the phase transition in In/Si(111) calculated from first principles~\cite{wall12}.
When $\mu$ moves even further toward the band edge, the uniform state becomes thermodynamically stable
while two local minima for $x\neq 0$ show that the dimerized states are metastable.
Finally, when $\mu$ approaches the band edge, we find a single-well potential 
indicating that the Peierls instability is suppressed. 
The variation of the grand canonical potential with $\mu$ explains the sensitivity of the transition
in indium wires to chemical doping~\cite{shim09,mori10,schm11,tera08,zhan14} and to optical excitations~\cite{wall12,tera08}. 
In particular, the observation that the uniform phase is stabilized in $n$-doped samples~\cite{tera08,zhan14} as well as by alkali 
adsorption-induced charge transfer~\cite{shim09,mori10} is naturally explained by the
occurrence of a metastable uniform state in the grand canonical potential in Fig.~\ref{fig:potential}.

If the temperature is raised without varying $\mu$, the grand canonical potential changes its shape progressively into
a single well but the uniform and dimerized states never exchange their relative energy 
positions~\cite{suppmat}.
Therefore, if we assume that $\mu$ deviates slightly from the middle of the gap, the uniform state is metastable at low temperature
but the structural transition remains continuous as in the canonical ensemble. 
Yet the actual electronic gap closes when one of the band edges reaches $\mu$ and thus
the metal-insulator transition occurs discontinuously and at a lower temperature than the structural transition.

In the 1D model, however, $\mu$ represents the influence of the substrate and thus it is a function of temperature rather than
an independent parameter.
(Equivalently, the dependence of the electron number on $\mu$ could change with
temperature~\cite{suppmat}.)
Moreover, a small change in $\mu$ is sufficient to change the shape of the grand canonical potential (see the inset of Fig.~\ref{fig:potential})
and thus to cause a discontinuous transition~\cite{suppmat}.
This scenario is compatible with recent first-principles simulations and experiments~\cite{hatt11,wall12,schm12,klas14,zhan14}.
Note that the dimerized configuration could be unstable toward the formation of domain walls (solitons)~\cite{zhan11a,kim12,cheo15}
but the study of spatial and thermal fluctuation effects is beyond the scope of this paper~\cite{heeg88,baeriswyl92,jeck94,jeck98b,horo75}.
The finding of a first order transition with a small reduction of the
order parameter in the critical region (see Fig.~6 in~\cite{suppmat})
justifies the neglect of fluctuations in first approximation.

The Peierls/CDW theory predicts the existence of collective excitations (amplitude modes)
which are Raman active~\cite{schu78,horo78,tuti91,gruener00}.
For the Peierls wave number $Q$ their frequency vanishes as $\omega(T) \propto \sqrt{\vert T-T_c \vert}$
when approaching $T_c$ in a continuous transition (phonon softening)~\cite{gruener00,tuti91}.
As the Peierls amplitude modes in In/Si(111) are essentially the shear and rotary modes, 
they should appear in the Raman spectrum at the $\Gamma$ point below $T_c$ and show significant (but incomplete) softening close 
to the first-order transition~\cite{suppmat}. 

\begin{figure}[tb]
\includegraphics[width=0.49\textwidth]{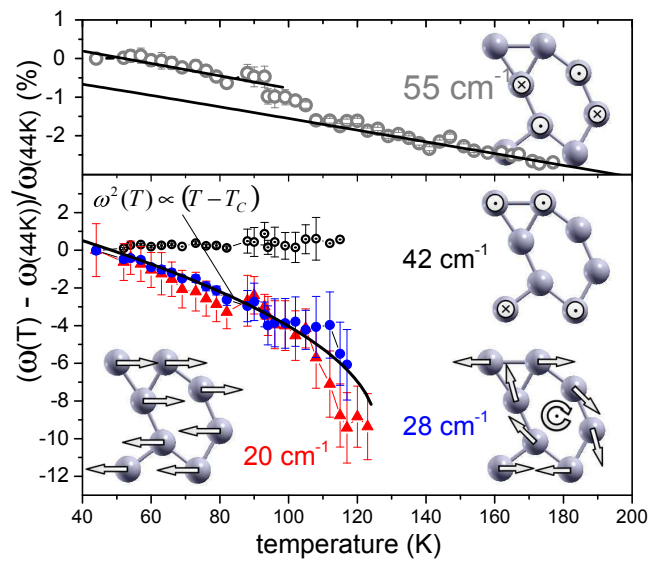}
\caption{\label{fig:phonons} (Color online) 
Temperature dependence of the normalized frequencies of Raman modes and sketches of the assigned eigenmodes.
The shear and rotary modes (red and blue symbols) at 20 and 28 cm$^{-1}$ are Peierls amplitude modes and exhibit a significant softening, 
while the mode at 42 cm$^{-1}$ (black symbols) remains at constant frequency and the one at 55 cm$^{-1}$ (grey symbols) shows only a moderate decrease
due to the lattice expansion. 
}
\end{figure}

Figure~\ref{fig:phonons} shows 
the temperature dependence of the normalized frequencies of some Raman spectra resonances measured for In/Si(111).
The resonances observed experimentally were assigned to specific vibrational modes by comparison to first-principles 
computations~\cite{flei03,flei07,spei10}.
Here we discuss the low-frequency modes at 20, 28, 42~cm$^{-1}$ in the ($8\times2$) phase and the 55~cm$^{-1}$ mode observed for both phases,
which all involve displacements of In atoms. 
The resonances at 20 and 28~cm$^{-1}$ (as measured at 44 K) are assigned to the shear and rotary modes. 
They exhibit a partial phonon softening when approaching the phase transition temperature and vanish above it. 
The mode at 42~cm$^{-1}$, in contrast, is at constant frequency with temperature while the mode at  
55~cm$^{-1}$ exhibits only moderate temperature shift. 
These observations agree qualitatively with our theoretical analysis but not with an order-disorder transition~\cite{gonz06,gonz09}.
The rotary and shear modes are strongly coupled to the CDW by the lateral displacements of the In atoms and
show the expected softening for Peierls amplitude modes, however, this softening remains only partial because
the transition is discontinuous. 
The 42 and 55 cm$^{-1}$ modes, in contrast, are related to vertical displacements of In atoms. Hence they are weakly 
coupled to the in-plane CDW and display a behavior related to the lattice expansion with temperature increase. 
Remarkably, the 42~cm$^{-1}$ mode shows no frequency shift at all, i.e. the lattice expansion is compensated by a stiffening of the involved In bonds. 
The 55~cm$^{-1}$ mode displays a side-effect drop in eigenfrequency at the phase transition.

In summary, we have shown that the transition observed in In/Si(111) is a grand canonical Peierls transition.
We think that the ongoing controversy about the nature of this transition can be solved 
by interpreting experiments and first-principles 
simulations~\cite{sni10,cho01,tani04,ahn04,tsay05,gonz05,gonz06,lope06,gonz09,wipp10,spei10,hatt11,schm12,wall12,kim13,klas14,zhan14}
within a grand canonical Peierls theory.  
In particular, it explains the observation of a metastable metallic phase at low temperature and the sensitivity of the
critical temperature to the substrate doping.
Grand canonical theories could explain other charge-donation related phenomena in atomic wires such as the reversible structural transitions 
in Au/Si(553) upon electron injection~\cite{pole13,pole14}.
The present work suggests that variations of the substrate-induced  chemical potential (e.g., with temperature or upon doping) 
is a key mechanism for understanding the realization of quasi-1D physics in atomic wires.

\begin{acknowledgments}
We thank S. Wippermann for helpful discussions. This work was done
as part of the Research Unit \textit{Metallic nanowires on the atomic scale: Electronic
and vibrational coupling in real world systems} (FOR1700) 
of the German Research Foundation (DFG) and was supported by
grants Nos.~JE~261/1-1.
Financial support by the \textit{Ministerium f\"ur Innovation, Wissenschaft und Forschung 
des Landes Nordrhein-Westfalen}, the \textit{Senatsverwaltung f\"ur Wirtschaft, Technologie und Forschung des Landes Berlin}, 
and the German \textit{Bundesministerium f\"ur Bildung und Forschung} is gratefully acknowledged.

\end{acknowledgments}

\end{document}